\documentstyle[12pt]{article}
\begin{document}
\begin{center}
{\large \bf Neutrino Tests of General and Special Relativity}\\
\vspace{1.0cm}
C. N. Leung\footnote{Talk presented at the Nufact'99 Workshop, 
Lyon, France, July 5-9, 1999.}\\
\bigskip
Department of Physics and Astronomy, University of Delaware\\ 
Newark, DE 19716, U.S.A.\\
\vspace{1.0cm}

{\bf Abstract}
\end{center}
We review the status of testing the principle of equivalence and 
Lorentz invariance from atmospheric and solar neutrino experiments.

\newpage
Let us begin by asking the question: do neutrinos obey the 
principle of equivalence?  The answer must of course come from 
experiments.  For ordinary matter, the most sensitive test of 
the equivalence principle comes from torsion balance experiments 
which measure the gravitational acceleration of macroscopic 
bodies.  The current best limit\cite{torex} is $10^{-12}$, i.e., 
all macroscopic bodies experience the same gravitational 
acceleration to an accuracy of one part in $10^{12}$.  Surely 
torsion balance experiments are not applicable to neutrinos.  
So the question is: what types of experiments are suitable for 
testing if neutrinos obey the principle of equivalence?

Suppose the neutrinos violate the principle of equivalence.  A 
consequence will be that different neutrino types (i.e., different
neutrino gravitational eigenstates) will couple to gravity with a 
slightly different strength.  Suppose the neutrino weak interaction 
eigenstates are not the same as their gravitational interaction 
eigenstates, but are linear superpositions of them.  Then, as a 
neutrino of definite flavour (e.g., $\nu_\mu$) travels through a 
gravitational field, the gravitational components of the flavour 
neutrino will develop different dynamical phases, which will result 
in neutrino flavour oscillations.  In other words, neutrino 
oscillation experiments provide a laboratory to test the 
equivalence principle for neutrinos\cite{Gasp,HL}. 

To parametrize the probabilities for neutrino oscillations arising 
from a violation of the equivalence principle (VEP), we follow the 
formalism in Ref.\cite{HLP} where the general equation governing 
neutrino flavour evolutions is found to be 

\begin{eqnarray}
i \frac{d}{dt} 
\left( \begin{array}{c} \nu_e \\ \nu_\mu \end{array} \right)
&=& 
\left\{
{\Delta m^2 \over 4E}
 \left[ \begin{array}{cc}
- \cos 2 \theta_M & \sin 2 \theta_M \\
\sin 2 \theta_M & \cos 2 \theta_M \end{array}
\right] \right. \nonumber \\
&+& \left.
E |\phi(r)| \Delta f
 \left[ \begin{array}{cc}
-\cos 2 \theta_G & e^{-i\beta} \sin 2 \theta_G \\
e^{i\beta} \sin 2 \theta_G & \cos 2 \theta_G \end{array}
\right] \right. \nonumber \\
&+& \left.
\frac{G_F N_e}{\sqrt{2}} 
\left[ \begin{array}{cc} 1 & 0 \\ 0 & -1 \end{array} \right]
\right\}
\left( \begin{array}{c} \nu_e \\ \nu_\mu \end{array} \right)_.
\label{evol}
\end{eqnarray} 

We have assumed two neutrino flavours ($\nu_e$ and $\nu_\mu$) 
for simplicity.  The first and third line on the right-hand 
side of Eq.(\ref{evol}) are familiar from the usual studies of 
neutrino oscillations.  The first line describes flavour mixing 
arising from nondegenerate neutrino vacuum masses\cite{Ponte}.  
Here $E$ denotes the neutrino's energy, $\Delta m^2 \equiv m_2^2 
- m_1^2$ is the neutrino mass-squared difference, and $\theta_M$ 
is the mixing angle between the weak interaction eigenstates 
and the mass eigenstates of the neutrinos.  The third line 
describes the matter effects on neutrino flavour 
evolutions\cite{W}.  Here $G_F$ is the Fermi constant and 
$N_e$ is the electron density in matter.  To study neutrino 
oscillations in vacuum, simply set $N_e = 0$.  The second line 
on the right-hand side of Eq.(\ref{evol}) describes neutrino 
flavour mixing arising from VEP.  Here $\phi(r)$, which 
satisfies the boundary condition $\phi(r \rightarrow \infty) 
\rightarrow 0$, is the gravitational potential through which 
the neutrinos propagate, $\theta_G$ is the mixing angle between 
the weak interaction eigenstates and the gravitational 
interaction eigenstates of the neutrinos, $\beta$ is a phase 
factor which comes about because we have mixing among three 
generally distinct sets of neutrino eigenstates: weak, 
gravitational, and mass.  Finally, $\Delta f$ is a measure of 
VEP defined as $\Delta f \equiv f_2 - f_1$, where $f_{1,2}$ 
are parameters introduced in the linearized neutrino-gravity 
interaction Lagrangian density\cite{BD},
\begin{equation}
 {\cal L}_{{\rm int}} = \frac{i f_j}{4} \sqrt{8 \pi G_N} 
 h^{\alpha\beta} [\bar{\nu_j} \gamma_\alpha \partial_\beta \nu_j 
 - (\partial_\alpha \bar{\nu_j}) \gamma_\beta \nu_j], ~~~~j = 1,2                        
\end{equation}
to gauge the deviation from universal coupling to gravity.  
Here $\nu_{1,2}$ denote the neutrino gravitational eigenstates, 
$h^{\alpha\beta}$ is the background gravitational field, and   
$G_N$ is the Newton gravitational constant.  Einstein's theory 
of general relativity predicts $f_1 = f_2 = 1$.  VEP occurs 
when $f_1 \neq f_2$.  

It can be concluded from Eq.(\ref{evol}) that 
\begin{enumerate}
 \item VEP leads to neutrino flavour oscillations even if the 
	neutrinos are massless or degenerate (in this case the 
	phase factor $\beta$ can be eliminated by redefinition 
	of the neutrino fields);
 \item neutrino oscillation tests of the principle of equivalence 
	involve two parameters: $\Delta f$ and $\theta_G$; it is 
	also necessary to know the value of the local gravitational 
	potential;
 \item VEP oscillations have a different energy dependence from 
	oscillations due to nondegenerate neutrino masses.
\end{enumerate}

The dependence of VEP oscillations on $\phi(r)$ is a consequence 
of violating the principle of equivalence.  It is also a source 
of uncertainty.  Table 1 in Ref.\cite{HLP} shows that our local 
value of $|\phi|$ spans from $6 \times 10^{-10}$ for the Earth's 
gravitational potential to about $3 \times 10^{-5}$ for the 
gravitational potential due to our local supercluster\cite{Ken}.  
Its value may be even bigger if the contributions from more distant 
sources can be estimated.  Since $\phi$ appears to be dominated by  
distant sources, its value varies little over a typical neutrino 
path length for earthbound experiments and may thus be treated 
as a constant.  It is then convenient to treat $|\phi| \Delta f$ 
as a single parameter in phenomenological analyses.

In the constant $\phi$ approximation, the VEP flavour transition 
probability has the familiar form ($\Delta m^2$ is assumed to 
be zero here):
\begin{equation}
 P(\nu_e \rightarrow \nu_\mu) = \sin^2 (2\theta_G) \sin^2 \left(
 \frac{\pi L}{\lambda_G} \right)_, 
\label{VEPprob}
\end{equation}
where $L$ is the neutrino path length and $\lambda_G$ is the 
oscillation length given by 
\begin{equation}
 \lambda_G = \frac{\pi}{E |\phi| \Delta f} 
  = 6.2~{\rm km} \Bigl(\frac{10^{-19}}{|\phi| \Delta f}
 \Bigr) \Bigl(\frac{1~{\rm GeV}}{E}\Bigr)_. 
\label{VEPlength}
\end{equation}
If we compare this with the well-known vacuum oscillation 
probability due to a neutrino mass difference, we see that the 
two cases are related by the formal connection, 
\begin{equation}
 \frac{\Delta m^2}{4E} \rightarrow E |\phi| \Delta f ~~~~~~
 {\rm and} ~~~~~~ \theta_M \rightarrow \theta_G.
\label{VEPsub}
\end{equation}
This connection can also be gleaned from Eq.(\ref{evol}).  
In contrast to the case of flavour oscillations induced by a 
neutrino mass difference where the oscillation length increases 
with the neutrino's energy, VEP oscillations are characterized 
by an oscillation length that decreases with increasing neutrino 
energy.  The two mechanisms may therefore be distinguished 
by measuring the neutrino energy spectrum\cite{PHL,BKL,LK,MN}.  
Note also that VEP oscillations will be more prominent in 
experiments with a large $E \cdot L$, i.e., high energy 
neutrinos and/or long path lengths, which is ideal for 
the neutrino factory experiments discussed in this workshop.

In a certain class of string theories, VEP may arise through 
interactions with a massless dilaton field\cite{DP}.  This type 
of equivalence principle violation may also lead to neutrino 
oscillations, but with an energy dependence which is the same 
as in oscillations due to nondegenerate neutrino masses\cite
{HLstring}.  It is therefore difficult to distinguish this type 
of VEP oscillations from the mass mixing oscillations, although 
one can still use data from oscillation experiments to constrain 
the relevant neutrino-dilaton couplings\cite{HLstring}.  We 
shall not consider further VEP oscillations from string theory, 
but will focus on the VEP oscillations from Eq.(\ref{evol}), 
which share the same distinctive energy dependence\cite{GHKLP} 
with neutrino oscillations arising from a possible breakdown of 
Lorentz invariance\cite{CG}.

If Lorentz invariance is violated, different (massless) neutrino 
species may have different maximum attainable speeds which are 
close to but not necessarily equal to $c$.  Neutrino oscillations 
can occur if the neutrino flavour eigenstates are linear 
superpositions of their velocity eigenstates, defined to be the 
energy eigenstates at infinite momentum, with the probability 
(for two-neutrino mixing)\cite{CG}
\begin{equation}
 P(\nu_e \rightarrow \nu_\mu) = \sin^2 (2\theta_v) \sin^2 \left(
 \frac{\pi L}{\lambda_v} \right)_, 
\label{VLIprob}
\end{equation}
where $L$ is the neutrino path length, $\theta_v$ is the mixing 
angle between the weak interaction eigenstates and the velocity 
eigenstates of the neutrinos, and $\lambda_v$ is the oscillation 
length given by 
\begin{equation}
 \lambda_v = \frac{2 \pi}{E \Delta v} 
 = 1.24~{\rm km} \Bigl(\frac{10^{-18}}{\Delta v}
 \Bigr) \Bigl(\frac{1~{\rm GeV}}{E}\Bigr)_. 
\label{VLIlength}
\end{equation} 
Here $\Delta v = v_2 - v_1$ is the difference between the 
speeds of the two neutrino velocity eigenstates.  Comparing 
Eqs.(\ref{VLIprob}) and (\ref{VLIlength}) with Eqs.(\ref{VEPprob}) 
and (\ref{VEPlength}), the similarities between the VLI 
(violation of Lorentz invariance) oscillations and VEP 
oscillations (for constant $\phi$) are obvious.  

Eqs.(\ref{VEPlength}) and (\ref{VLIlength}) indicate the 
sensitivity we can expect for VEP and VLI tests from neutrino 
oscillation experiments.  Assuming maximal mixing, there will be 
appreciable flavour transitions if $\pi L/\lambda_{G,v} 
\sim O(1)$.  For atmospheric neutrinos, $E \sim (0.1 - 10^3)$ 
GeV and $L \sim (20 - 10^4)$ km, hence $|\phi| \Delta f$ (or 
$\Delta v/2$) in the range $(10^{-26} - 10^{-19})$ can be 
probed.  For solar neutrinos, $E \sim (0.1 - 10)$ MeV and, for 
vacuum oscillations, $L \sim 10^8$ km, hence $|\phi| \Delta f$ 
(or $\Delta v/2$) in the range $(10^{-25} - 10^{-23})$ can be 
probed.  For MSW solution to the solar neutrino problem, the 
solar gravitational potential, which is about $1.7 \times 
10^{-5}$ at the solar core and decreases to about $ 2 \times 
10^{-6}$ at the surface of the Sun, plays a significant role 
and the constant $\phi$ approximation is not as good here.  
Nevertheless we can obtain an order-of-magnitude estimate 
for the sensitivity limit : $|\phi| \Delta f$ (or $\Delta v/2$)
$\sim (10^{-23} - 10^{-22})$, by letting $L = 7 \times 
10^5$ km, the mean solar radius (see Fig. 2 in Ref.\cite{BKL} 
for a more accurate estimate).  For long-baseline experiments 
such as those envisaged for future neutrino factories, 
$E \sim (10 \, - \, 100)$ GeV and $L \sim ({\rm few} \times 10^2 
\, - \, {\rm few} \times 10^3)$ km, hence a sensitivity of 
$(10^{-25} \, - \, 10^{-23})$ for $|\phi| \Delta f$ (or 
$\Delta v/2$) can be reached.  These estimates demonstrate that 
neutrino oscillations can provide a very sensitive test for the 
fundamental principles of general and special relativity.

We now confront VEP and VLI with experiments.  Because of the 
limited time, we shall consider only atmospheric and solar 
neutrino experiments.  

Positive evidence for $\nu_\mu \rightarrow \nu_\tau$ 
transitions has been established by the Super-Kamiokande 
(SK) Collaboration\cite{SKatmos}.  An analysis\cite{FLY} of 
the 535 days of SK data on sub-GeV and multi-GeV events 
found that VEP and VLI oscillations were consistent with 
the data, although the $\chi^2$ fit was not as good as 
the fit for mass mixing oscillations.  However, later 
analyses\cite{LL,FLMS} that included more recent SK 
data\cite{SKupmu} on upward-going muon events, which 
corresponded to atmospheric neutrinos with higher energies: 
$E$ up to $\sim 10^3$ GeV, found that VEP and VLI oscillations 
were not compatible with the data.  In particular, Fogli 
{\it et al.}\cite{FLMS}, did a very thorough analysis of the 
VEP and VLI mechanisms with the SK data.  They allowed a 
power-law energy dependence for the oscillation length: 
$\lambda \propto E^{-n}$, and found that, at 90\% C.L., 
$n = - 0.9 \pm 0.4$, which identified the mass mixing mechanism 
as the mechanism for the observed $\nu_\mu \rightarrow \nu_\tau$ 
transitions.  VEP and VLI oscillations, which correspond to 
$n = 1$, are clearly incompatible with the data.  The possibility 
of both mass mixing and VEP (or VLI) mechanisms contributing to 
the flavour transitions, as described in Eq.(\ref{evol}), was 
also studied in Ref.\cite{FLMS}.  It was found that, for a wide 
range of parameter choices, including the VEP (or VLI) 
contributions did not improve the fit for the mass mixing 
mechanism, which allowed the authors to obtain the 90\% C.L. 
upper bounds for the $\nu_\mu$-$\nu_\tau$ sector:
\begin{equation}
 |\Delta f| < 10^{-19}, ~~~~~~~~
 |\Delta v| < 6 \times 10^{-24},
\end{equation}
independent of the values of the corresponding mixing angles.  
The limit on $\Delta f$ is obtained by assuming $|\phi| = 3 
\times 10^{-5}$, the contribution from the local supercluster.  
This limit on VEP is seven orders of magnitude more stringent 
than the best limit from torsion balance experiments\cite{torex}.  
The limit on $\Delta v$ is also the most stringent limit on VLI 
to-date.  Note finally that these bounds are consistent with the 
range of sensitivity discussed above and are within about an  
order of magnitude of the sensitivity limit for long-baseline 
experiments. 

As constraining as the SK atmospheric neutrino data are, they 
only tell us about VEP and VLI in the $\nu_\mu$-$\nu_\tau$ sector.  
To study flavour transitions involving $\nu_e$, we turn to solar 
neutrino experiments.  The observed solar neutrino deficit may be 
a result of long-wavelength neutrino vacuum oscillations or 
matter-enhanced flavour transitions (MSW effect)\cite{W,MS} in 
the Sun.  Based on the Standard Solar Model predictions in 
Ref.\cite{BP92}, it was found in Ref.\cite{PHL} that only 
matter-enhanced VEP oscillations were compatible with the 
available solar neutrino data then.  However, using the improved 
Standard Solar Model predictions\cite{BP98} and the latest solar 
neutrino data\cite{Home98,SAGE99,GALLEX99,SKsun99}, and treating 
the $^8$B neutrino flux as a free parameter, a recent 
study\cite{GNF} finds that long-wavelength VEP oscillations 
can also account for the solar neutrino deficit.  The authors 
of Ref.\cite{GNF} examined constraints from the observed solar 
neutrino rates and from the SK spectral data.  They found that, 
for $\nu_e \rightarrow \nu_\tau$ transitions, the allowed 
parameter space lay in the region 
\begin{equation}
 0.65 \leq \sin^2(2\theta_G) \leq 1.0 ~~{\rm and}~~ 
 3.3 \times 10^{-20} < |\Delta f| \leq 3.3 \times 10^{-18} 
\label{VEPvac}
\end{equation}
for VEP and 
\begin{equation}
 0.65 \leq \sin^2(2\theta_v) \leq 1.0 ~~{\rm and}~~ 
 2.0 \times 10^{-24} < |\Delta v| \leq 2.0 \times 10^{-22} 
\label{VLIvac}
\end{equation}
for VLI.  We have again used $|\phi| = 3 \times 10^{-5}$ to 
obtain the limits on $\Delta f$ in (\ref{VEPvac}).  For 
$\nu_e \rightarrow \nu_\mu$ transitions, constraints\cite{PKM} 
from the recent accelerator neutrino data from the CCFR 
Collaboration\cite{CCFR} reduce the allowed range for $|\Delta f|$ 
and $|\Delta v|$ to $3.3 \times 10^{-20} < |\Delta f| < 
6.7 \times 10^{-19}$ and $2.0 \times 10^{-24} < |\Delta v| < 
4.0 \times 10^{-23}$, for the same range of values for the 
corresponding mixing angle.  

The VEP and VLI mechanisms also admit a MSW solution to the 
solar neutrino problem.  One generally finds two disjointed 
allowed regions, one for small mixing angles and one for large 
mixings\cite{HLP,PHL,BKL,PKM,MK}.  Figure 1, which is due to 
P. I. Krastev, shows the 99\% confidence level allowed regions 
for $|\Delta f|$ and $\sin^2(2\theta_G)$ derived from comparing 
all available data on the solar neutrino rates, including the 
825 days of SK data\cite{SKsun99}, with the Standard Solar Model 
predictions in Ref.\cite{BP98}.  Figure 1 is obtained by using 
only the solar gravitational potential for $\phi(r)$ in 
Eq.(\ref{evol}) to determine the neutrino flavour evolution.  
For the $\nu_e \rightarrow \nu_\mu$ channel, the allowed regions  
in Figure 1 are completely ruled out by the constraints from the 
CCFR data\cite{PKM}.  For the $\nu_e \rightarrow \nu_\tau$ 
channel, only the large mixing angle region is incompatible 
with the CCFR data\footnote{The authors of Ref.\cite{PKM} claim 
that a three-neutrino mixing analysis rules out the MSW 
allowed regions completely, for both $\nu_e \rightarrow \nu_\mu$ 
and $\nu_e \rightarrow \nu_\tau$ transitions.}.  The remaining 
small mixing region may be constrained by the SK data on the 
recoil-electron energy spectrum, although an analysis\cite{MK} 
based on earlier (504 days) SK data did not find such a 
constraint.  

In conclusion, {\em in addition to} helping determine the 
masses and mixing angles of neutrinos, neutrino oscillation 
experiments constitute a very useful tool for testing the 
principle of equivalence and Lorentz invariance.  Current 
solar and atmospheric neutrino data already provide very 
stringent limits.  For the $\nu_\mu$-$\nu_\tau$ sector, 
atmospheric neutrino data imply that the principle of 
equivalence cannot be violated by more than one part in 
$10^{19}$ and Lorentz invariance cannot be violated by more 
than $6 \times 10^{-24}$.  Solar neutrino data provide limits 
for both the $\nu_e$-$\nu_\mu$ and $\nu_e$-$\nu_\tau$ sectors.  
The exact limits depend on the assumption of either MSW 
oscillations or long-wavelength vacuum oscillations, with the 
long-wavelength solution imposing the stronger bounds.  Future 
long-baseline experiments can help improve these limits with 
sufficiently long (thousands of kilometers) baselines and 
sufficiently high energy (hundreds of GeV) neutrinos.  A 
neutrino factory will be a big access for reaching these 
goals. \\

\begin{center}
{\bf Acknowledgement}\\
\end{center}
\smallskip
This work was supported in part by the U.S. Department of Energy 
under grant DE-FG02-84ER40163. \\


\bigskip


\begin{thebibliography}{999}

\bibitem{torex} 
V. B. Braginsky and V. I. Panov, {\em Sov. Phys. JETP} {\bf 34} 
(1972) 463.  For a more recent experiment, see, e.g., B. R. 
Heckel, {\em et al.}, {\em Phys. Rev. Lett.} {\bf 63} (1989) 2705.

\bibitem{Gasp} 
M. Gasperini, {\em Phys. Rev. D} {\bf 38} (1988) 2635; {\em Phys. 
Rev. D} {\bf39} (1989) 3606. 
 
\bibitem{HL}
A. Halprin and C. N. Leung, {\em Phys. Rev. Lett.} {\bf 67} (1991) 
1833.

\bibitem{HLP}
A. Halprin, C. N. Leung, and J. Pantaleone, {\em Phys. Rev. D} 
{\bf 53} (1996) 5365.

\bibitem{Ponte}
B. M. Pontecorvo, {\em Zh. Eksp. Teor. Fiz.} {\bf 34} (1958) 
247 [{\em Sov. Phys. JETP} {\bf 7} (1958) 172]; Z. Maki, 
M. Nakagawa, and S. Sakata, {\em Prog. Theor. Phys.} {\bf 28} 
(1962) 870.

\bibitem{W}
L. Wolfenstein, {\em Phys. Rev. D} {\bf 17} (1978) 2369; 
{\em Phys. Rev. D} {\bf 20} (1979) 2634.

\bibitem{BD}
D. G. Boulware and S. Deser, {\em Ann. Phys.} (N.Y.) {\bf 89} 
(1975) 193.

\bibitem{Ken}
I. R. Kenyon, {\em Phys. Lett. B} {\bf 237} (1990) 274.

\bibitem{PHL}
J. Pantaleone, A. Halprin, and C. N. Leung, {\em Phys. Rev. D} 
{\bf 47} (1993) R4199.

\bibitem{BKL}
J. N. Bahcall, P. I. Krastev, and C. N. Leung, {\em Phys. Rev. D} 
{\bf 52} (1995) 1770. 

\bibitem{LK}
C. N. Leung and P. I. Krastev, in {\em The Albuquerque Meeting}, 
Proceedings of the 8th Meeting of the Division of Particles and 
Fields of the American Physical Society, ed. S. Seidel (World 
Scientific, Singapore, 1995), p.1544. 

\bibitem{MN}
H. Minakata and H. Nunokawa, {\em Phys. Rev. D} {\bf 51} (1995) 
6625.

\bibitem{DP}
T. Damour and A. M. Polyakov, {\em Gen. Rel. Grav.} {\bf 26} 
(1994) 1171; {\em Nucl. Phys.} {\bf B423} (1994) 532.

\bibitem{HLstring}
A. Halprin and C. N. Leung, {\em Phys. Lett. B} {\bf 416} (1998) 
361.

\bibitem{GHKLP}
S. L. Glashow, A. Halprin, P. I. Krastev, C. N. Leung, and 
J. Pantaleone, {\em Phys. Rev. D} {\bf 56} (1997) 2433. 

\bibitem{CG}
S. Coleman and S. L. Glashow, {\em Phys. Lett. B} {\bf 405} 
(1997) 249.

\bibitem{SKatmos}
Super-Kamiokande Collaboration, Y. Fukuda {\it et al}., 
{\em Phys. Rev. Lett.} {\bf 81} (1998) 1562; {\em Phys. 
Lett. B} {\bf 433} (1998) 9; {\em Phys. Lett. B} {\bf 436} 
(1998) 33.

\bibitem{FLY}
R. Foot, C. N. Leung, and O. Yasuda, {\em Phys. Lett. B} 
{\bf 443} (1998) 185.

\bibitem{LL}
P. Lipari and M. Lusignoli, {\em Phys. Rev. D} {\bf 60} (1999) 
013003.

\bibitem{FLMS}
G. L. Fogli, E. Lisi, A. Marrone, and G. Scioscia, {\em Phys. 
Rev. D} {\bf 60} (1999) 053006. 

\bibitem{SKupmu}
Super-Kamiokande Collaboration, Y. Fukuda {\it et al}., 
{\em Phys. Rev. Lett.} {\bf 82} (1999) 2644. 

\bibitem{MS}
S. P. Mikheyev and A. Yu Smirnov, {\em Yad. Fiz.} {\bf 42} 
(1985) 1441 [{\em Sov. J. Nucl. Phys.} {\bf 42} (1985) 913].

\bibitem{BP92}
J. N. Bahcall and R. K. Ulrich, {\em Rev. Mod. Phys.} {\bf 60} 
(1988) 297; J. N. Bahcall and M. H. Pinsonneault, {\em Rev. Mod. 
Phys.} {\bf 64} (1992) 885.

\bibitem{BP98}
J. N. Bahcall, S. Basu, and M. H. Pinsonneault, {\em Phys. Lett. 
B} {\bf 433} (1998) 1.

\bibitem{Home98}
Homestake Collaboration, K. Lande {\it et al}., {\em Astrophys. J.} 
{\bf 496} (1998) 505.

\bibitem{SAGE99}
SAGE Collaboration, J. N. Abdurashitov {\it et al}., 
astro-ph/9907113.

\bibitem{GALLEX99}
GALLEX Collaboration, W. Hampel {\it et al}., {\em Phys. Lett. B} 
{\bf 447} (1999) 127.

\bibitem{SKsun99}
Y. Suzuki for the Super-Kamiokande Collaboration, talk at the 
Lepton-Photon '99 Conference.

\bibitem{GNF}
A. M. Gago, H. Nunokawa, and R. Zukanovich Funchal, hep-ph/9909250.

\bibitem{PKM}
J. Pantaleone, T. K. Kuo, and S. W. Mansour, hep-ph/9907478.

\bibitem{CCFR}
CCFR Collaboration, A. Romosan {\it et al}., {\em Phys. Rev. 
Lett.} {\bf 78} (1997) 2912; D. Naples {\it et al}., {\em Phys. 
Rev. D} {\bf 59} (1999) 031101. 

\bibitem{MK}
S. W. Mansour and T. K. Kuo, {\em Phys. Rev. D} {\bf 60} (1999) 
097301.

\end{thebibliography}
\end{document}